\documentclass[amsmath,amssymb]{revtex4}

\usepackage{bm}
\usepackage{color}
\usepackage{dcolumn}
\usepackage{epsfig}
\usepackage{graphicx,rotating}
\usepackage{longtable, booktabs}
\usepackage{mhchem}
\usepackage{gensymb}
\usepackage{multirow}
\usepackage{balance}
\usepackage[caption=false]{subfig}

\newcolumntype{C}[1]{>{\centering\arraybackslash}m{#1}} 

\begin{document}

\title{Role of Inorganic Cations in the Excitonic Properties of Lead Halide Perovskites}

\author{Ma\l{}gorzata Wierzbowska}
\affiliation{Institute of High Pressure Physics, Polish Academy of Sciences, Soko\l{}owska 29/37, 01-142 Warsaw, Poland}
\author{Juan J. Mel\'endez}
\affiliation{Department of Physics, University of Extremadura. Avda. de Elvas, s/n, 06006, Badajoz, Spain}
\affiliation{Institute for Advanced Scientific Computing of Extremadura, Badajoz, Spain}

\date{\today}

\begin{abstract}
We theoretically investigate lead iodide perovskites of general formula APbI$_3$ for a series of metallic cations 
(namely Cs$^+$, Rb$^+$, K$^+$, Na$^+$ and Li$^+$) 
by means of the density functional theory, GW method and Bethe-Salpeter equation including 
spin-orbit coupling. 
We demonstrate that the low-energy edges (up to 1.3 eV) of the absorption spectra are dominated by 
weakly bound excitons, with binding energies $E_b$ $\sim$ 30-80~meV,
and the corresponding intensities grow as metallic 
cations become lighter. 
The middle parts of the spectra (1.8-2.4 eV), on the other hand, contain optical dipole transitions comprising  
more strongly bound excitons ($E_b$ $\sim$ 150-200~meV) located at PbI$_3$. 
These parts of the spectra correspond to the optical-gain wavelengths which are experimentally achieved 
in optically-pumped perovskite lasers. 
Finally, the higher energy parts, from about 2.8~eV (LiPbI$_3$) to 4.3 eV (CsPbI$_3$), 
contain optical transitions with very strongly bound excitons ($E_b$ $\sim$ 220-290~meV)
located at the halide atoms and the empty states of the metallic cations. 
\end{abstract}

\maketitle

\section{Introduction}

Lead halide perovskites, with general chemical formula APbX$_3$, where A and X stand for a cation and a halide anion, 
respectively, are members of a family of compounds which has revolutionized the field of Optoelectronics 
since the first decade of this century \cite{Kojima09}. 
Chemical composition, fabrication routes or morphological design are factors which provide lead halide perovskites 
with an exceptional set of physico-chemical properties, including large absorption coefficient, 
long diffusion lengths for bipolar charge transport, high photoluminiscence quantum yield, 
low defect state densities or narrow linewidths, which makes them to find applications as components of solar cells, 
photodetectors, light emitting devices and optically pumped lasers, among others \cite{Zeng19, R2021}. 

In particular, organic-inorganic hybrid perovskites (HPs) have deserved considerable interest, mainly in relation 
to their use as components of high-performance solar cells \cite{Ansari18, Dedecker20, Leng20, Yang20, Marimuthu22}. 
For these perovskites, the A cation is typically methylammonium [CH$_3$NH$_3^+$, MA] or formamidinium [CH(NH$_2$)$_2^+$, FA], 
and doping species are often added to improve stability of performance. In this context, considerable advances have been reached 
in terms of efficiency of HPs solar cells, which has evolved from the preliminar 3.8 \% \cite{Kojima09} in 2009 to 
the maximum value of 25.6 \% reported so far \cite{Jeong21}. 
As a reference, the efficiency of first-generation silicon cells is around 30 \%. 
Considering that HPs are relatively cheap to synthesize, one may be tempted to think that they are optimal candidates 
to replace the costly Si-based solar cells. Unfortunately, the current situation is not so ideal, 
and the use of HPs in commercial devices is severely complicated by the presence of toxic metals and by structural instability 
issues \cite{Dedecker20, Schileo20}. Briefly, these are related to the exposure to environmental humidity 
(which irreversibly decomposes the HP onto its precursors and volatile acid compounds), \cite{Huang17} 
heat (decomposition occurs for some perovskites at temperatures as low as 40~$^{\circ}$C) \cite{Juarez18} 
and oxidation in presence of light, even for very low oxygen levels \cite{Aristidou17,Bryant16}, among others. 
Several strategies are being adopted to improve HPs stability, including standard encapsulation, tailoring composition 
to enhance tolerance to oxidation and doping, but none of them ensures the reliability of a HP-based solar cells 
at working conditions during long time \cite{Schileo20}. 

Degradation of HPs is partially due to the fact that the MA and FA organic cations are volatile. 
Structural stability may be then improved by replacing these cations by inorganic (typically alkaline) 
ones \cite{Liu18, Zhang18, Yang18, Sutton16}. Apart from an enhanced stability  in many cases, all-inorganic perovskites (IPs) 
exhibit interesting optoelectronic properties including long diffusion lengths and lifetimes for recombination of 
charge carriers \cite{Locardi18, Shi15, Wehrenfennig14} and strong optical absorption, \cite{Wolf14} which makes them 
promising components not only of solar cells, but also for light-emitting diodes (LEDs) and 
lasers \cite{Li17, Shan17, Evans18}. By varying the A cation, as well as the halide (Cl, Br, I) and B-site metal 
(typically Pb, Sn or Ge), one gets virtually a myriad of IPs, most of them still not well (or not at all) characterized. 
The archetypical IP is the CsPb(Br$_x$I$_{1-x}$)$_3$ system, whose state-of-art may be found elsewhere \cite{Zeng19}. 
In the context of this work, the following comments are pertinent:
\begin{enumerate}
	\item 
Despite the aforementioned enhanced structural stability of IPs, they display a rich phase transition phenomenology 
which affects their performance. For instance, CsPbI$_3$ crystallizes in the the cubic $\alpha$-phase 
(with a gap of 1.7~eV) above 593~K, and decomposes onto orthorhombic phases below room temperature \cite{Liu2020}.
Furthermore, there is a severe decomposition of the $\beta$ and $\gamma$ orthorhombic phases to the photovoltaically-inactive non-perovskite 
$\delta$-phase in presence of water vapor \cite{Becker19,exp2,exp3}.
The origin of this decomposition is the relatively small radius of Cs$^+$, which does not suffice to stabilize the PbI$_6$ octahedra 
of the perovskite structure. Despite the phase transformation is reversible upon annealing, the optoelectronic properties of 
the non-perovskite structure are poor. Strategies such as microstructural or dimensional control, doping or molecular modification 
have been used with the aim to improve the power conversion efficiency (PCE) of the CsPbI$_3$-based solar cells, 
but this remains inadequate for efficient photovoltaic devices \cite{Zeng19}. 
Bulk CsPbBr$_3$ is stable in an orthorhombic perovskite phase at room temperature, but its gap ($\sim$ 2.3~eV) is too high 
for photovoltaic applications \cite{Sebastian15}. The mixed halides CsPb(Br$_x$I$_{1-x}$)$_3$, on the other hand, 
display gaps between 1.7 and 2.3~eV, which overlap the optimum range for photovoltaic devices, and are stable for some 
compositions \cite{Nasstrom20}. The most widely studied member of this group is perhaps CsPbI$_2$Br, whose gap ranges 
between 1.8 and 1.9~eV. This makes it a prominent candidate for solar cells applications, although the highest PCE 
is still well below that for HPs. Moreover, a very short exciton lifetime in CsPbI$_2$Br
makes it also a promising material for LEDs and lasers.\cite{ultrafast}
	\item Regarding mixed-halide perovskites, several approaches have been adopted \cite{Zeng19}. 
For the purpose of this work, we will limit ourselves to chemical modification. Noticeable structural stabilization and 
increased PCEs from $\sim$ 4 \% of undoped CsPbI$_2$Br have been achieved by incorporating cations such as 
Ge$^{2+}$, Sn$^{2+}$, Sr$^{2+}$, Bi$^{3+}$, Ca$^{2+}$, Mn$^{2+}$, Sb$^{3+}$, Ni$^{2+}$ or lanthanides 
La$^{2+}$, Sm$^{2+}$ and Eu$^{2+}$ \cite{Yang18, Liang17, Lau17, Hu17, Lau18, Bai18, Xiang18, Chen19, Chen20, Patil20, Mali21} and, 
especially, halides such as SrCl$_2$ (PCE = 16.07 \%~),\cite{Qiao21} GdCl$_3$ (PCE = 16.24 \%~),\cite{Pu22} FeCl$_2$ 
(PCE = 17.1~\%~)\cite{Ozturk21} or InCl$_3$ (PCE = 17.45 \%~)\cite{Mali21}. All these cations incorporate to the perovskite 
structure at the Pb-site, which enhances structural stability because the doping cation shrinks the octahedral perovskite voids 
and reduces the lattice energy. The effect on optical and transport properties is related to the fact that 
the bottommost conduction band arises mainly from the Pb $p$-orbitals. Possible candidates for the A-cation, apart from Cs$^{2+}$, 
have been much scarcely considered, despite early theoretical works point out that the bandgap of metal-halide perovskites depends on
 the metal-halide-metal bond angle and, therefore, the bandgap could be tuned by adequate choices of the A cations \cite{Filip14}. 
In this context, successful fabrication of Rb-based perovskites was reported as nanocrystals \cite{Rb1,Rb2}, 
although Rb (as also K) usually acts as a co-dopant \cite{Science,Science2,Rb2}. Even smaller cations such as Na$^+$ could be 
incorporated to double perovskites as well, like in Cs$_2$NaBiCl$_6$ and Cs$_2$NaSbCl$_6$, but they locate at the B-sites 
(occupied by Pb in single perovskites) in order to alternate and balance the trivalent cations (Bi$^{3+}$, Sb$^{3+}$). \cite{d1,d2,d3,d4}	
	\item 
From the point of view of applications, perovskites are used in photovoltaic or optoelectronic devices in such a way that, 
at working conditions, different regions of their optical spectrum are excited. In particular, this means 
that IPs perovskites must work under diverse exciton binding energy ($E_b$) conditions. 
For instance, their usage as parts of photovoltaic devices (such as solar cells) is affected by material chemical 
(crystallization control during fabrication, defect and interface engineering, etc. \cite{Zeng19}) as also physical factors, 
namely a suitable bandgap as well as weakly bound excitons, easily dissociable into free charge carriers, 
to form at the onset of the optical absorption spectrum. Recent advances on perovskite solar cells have been reviewed elsewhere 
\cite{Suresh21,Yan22}. On the other hand, their use in light-emitting devices require again a proper bandgap as well as the 
existence of more strongly-bound (thus more stable) excitons, 
since these ease efficient recombination even at low carrier densities \cite{devices}. 
In this respect, despite a number of papers deal with advances on laser devices with perovskites as optical-gain materials 
\cite{R2021,R1, R2, R3, R4, R5}, there is much room for performance enhancement, especially for  optically pumped lasers. 
These devices may be classified according to their chemical structure, morphology and dimensionality, as well as to the lasing mode, 
emission wavelength and energy density threshold for the optical excitation. \cite{R2021} 
Bulk, quasi-2D structure and nanocrystals were reported to work under pumping thresholds ranging from 0.25 to 420~$\mu$J/cm$^2$, 
with emission wavelengths reported between 400 and 790~nm (1.56 - 3.1~eV). The common property of all this family of devices is 
the excitation energy for laser pumping (2.2 - 3.5~eV), which is well above the conduction band minimum in each case. 
This contrasts with electrically pumped lasers, for which the driven excitations locate near the fundamental bandgap -incidentally, there there is still a quest to construct first such a device with either IPs or HPs.
Therefore, perovskites entering into components of lasers must fulfill different requirements for optimal performance depending on 
the pumping mechanism: optically-pumped lasers require strongly bound excitons to be located below the pumping energy in 
the optical absorption spectrum, whereas electrically-pumped ones require excitons to form at the absorption edge.	
\end{enumerate}

With these ideas in mind, the main goal of this paper is to study the effect of the central A$^+$ cation between the inorganic 
PbI$_6$ octahedra in the all-inorganic perovskite APbI$_3$ structure. In particular, we are first aimed to investigate 
which energetic position of the A$^+$ states, and which transitions between the later and PbI$_3$ states, could be interesting 
for lasing applications. Just a few experimental works, and even less theoretical studies, focus on the role of inorganic A$^+$ cations. 
The most popular among experimentalists is Cs$^+$, reported as the more effective element for enhancing photostability 
when incorporated as a co-dopant in FA-based HPs\cite{Cs-1,Cs-2,triple}. Pure Cs-based perovskite lasers have been constructed 
as well\cite{C1,C2,C3,C4,C5}, but no attention has been paid to other doping species which, according to the previous discussion, 
enhance stabilization and modify the bandgap of IPs. 

Secondly, we are aimed to investigate the possible formation of excitons in these perovskite systems, not only close to the absorption edge, but also at much higher excitation energies. This interest arises from the fact that high-energy regions of the absorption spectrum could be achieved during the operation of pumped lasers. As will be discussed below, inorganic cations considered herein yield a single empty state within the conduction manifold up to 5 eV above the bandgap. This empty state is well localized (i.e., not hybridized with the lead-halide frame) at the A$^+$ cation, and it gives rise to the formation of bound excitons. By contrast, organic cations yield many empty states, which obscures the theoretical analysis. Incidentally, we have focused on inorganic cations because, unlike organic ones, they do not rotate. Rotations of the relatively large organic cations cause large distorsions of the inorganic frame \cite{breath,Rot-2,Rot-3} and switches the gap to indirect  \cite{Hutter-nature}, both effects making difficult any study of optical properties. 

Thirdly, we will be interested in establishing the nature of optical transitions in terms of the electron and hole states and 
their locations within the bands structure, since this could settle the basis for future developments. 
Experimental detection of the optical absorption spectra does not enable to distinguish between electron excitations purely 
localized at the PbI$_3$ states and those involving also empty states of the A$^+$ cation. 
Theoretical tools become then indispensable to examine the origin of the optical dipole transitions. 
Typically, theoretical studies adopt \textit{ab initio} many-body perturbation theory (MBPT) schemes, 
namely the non-selfconsistent GW and the Bethe-Salpeter equation (BSE) methods \cite{mbpt,bse}, 
built on top of the Bloch functions, which are spinors if spin-orbit coupling (SOC) is included, 
and computed within the density functional theory (DFT) formalism. 
This is the approach used in this work, where the absorption spectra of lead iodide perovskites APbI$_3$, 
with A = Cs, Rb, K, Na and Li, are examined up to 5 eV for the bands contribution to the optical dipole transitions 
and the corresponding exciton binding energies. The first three compounds give rise to stable structures, 
whereas the later two, too light to yield stability, must be considered as "theoretical limits" in this context. 
Special attention is paid to the selection of the excitations from iodide to A$^+$ cations from those which move an electron from I to Pb.

\section{Theoretical details}

\begin{figure}[htb!]
	\centering
	\includegraphics[width=0.6\columnwidth]{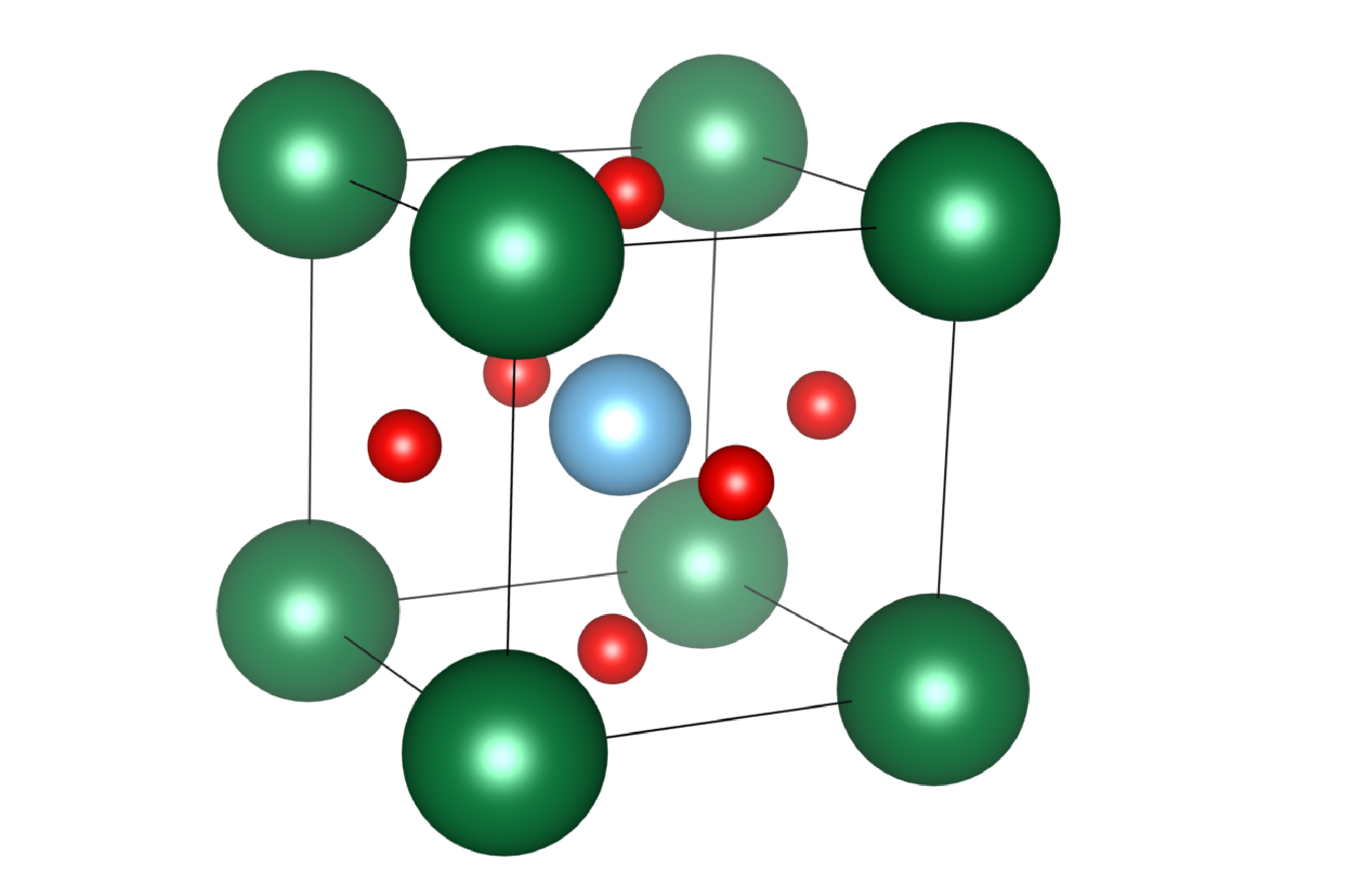}
	\caption{Archetype of the perovskite structure APbI$_3$, with 
	Pb at the origin (green), A$^+$ cation at the center of the cube (blue)
	and halide atoms at the cell-wall centers (red).}
	\label{F0}
\end{figure}

Lead halide perovskites occur in the cubic, tetragonal and orthorhombic crystal structures, 
depending on chemical composition and temperature.\cite{Nasstrom20}  
In this work, we have considered the simple cubic structure;
Fig.~\ref{F0} depicts the archetypical perovskite structure.
For each choice of the A$^+$ cation, the lattice parameter was set to the experimental value for 
CsPbI$_3$ (6.297~\AA \cite{Liu2020}) and subsequently optimized by the Broyden-Fletcher-Goldfarb-Shanno algorithm. 

DFT calculations including spin-orbit coupling (DFT+SOC) 
were performed under the generalized-gradient approximation (GGA)
using the Quantum ESPRESSO plane-wave package.\cite{qe} For completeness, a set of calculations under 
DFT without SOC was carried out as well. 
The atomic cores were described with Martin-Troulier norm-conserving full-relativistic pseudopotentials 
using the Perdew-Burke-Ernzerhof 
parametrization of the exchange-correlation functional. 
A plane-wave energy cutoff of 100 Ry was used, and the Brillouin zone (BZ) was sampled with 
the 6$\times$6$\times$6 Monkhorst-Pack mesh.\cite{mesh} 
The DFT band structures were interpolated from maximally-localized Wannier functions (MLWFs)\cite{wan1,wan2} 
using the Wannier90 code \cite{wan90} and plotted along high-symmetry lines, 
namely $\Gamma(0,0,0) \rightarrow R\left(\frac 12,\frac 12, \frac 12\right) \rightarrow 
X\left(\frac 12,0,0\right)\rightarrow M\left(\frac 12,\frac 12,0\right) \rightarrow \Gamma \rightarrow X$.

Non-selfconsistent GW and BSE calculations were performed on top of the DFT+SOC solutions 
using the Yambo code.\cite{yambo1,yambo2} 
GW calculations were carried out under the plasmon-pole approximation\cite{plaspol}. 
The plane-waves energy cutoffs were set to 30~Ry for the Coulomb and exchange interactions and to 6~Ry 
for the response block size, 
and 1000~bands were included in the response and GW summations. The BZ sampling for the GW was 4$\times$4$\times$4. 
As for the bare DFT+SOC, the GW energies were transferred to the Wannier90 code for bands interpolation 
and projection onto the A$^+$ cation localized MLWFs. 

The BSE was solved in two ways: 1) using a very dense 18$\times$18$\times$18 Monkhorst-Pack mesh 
to sample the BZ, in order to obtain 
the convergent absorption spectrum within the Haydock solution scheme, and 2) using a moderate 8$\times$8$\times$8 mesh for 
the full-diagonalization scheme, which enables the analysis of the dipole optical transitions 
in terms of interband excitations at specific $k$-points. 
For BSE, 200~bands were summed to build up the polarization function, while 36~bands were occupied. 
The plane waves up to energies of 10~Ry for the exchange components and 
4~Ry for the screening and response block-size were used. 
The electron-hole ($e-h$) pairs were formed by the 6~occupied and 8~unoccupied bands.

\section{Results and discussion}
\subsection{Band structure}
\begin{figure*}[htb!]
\begin{tabular}{ccc}
\hspace{-1.0cm} \includegraphics[scale=0.7,angle=0.0]{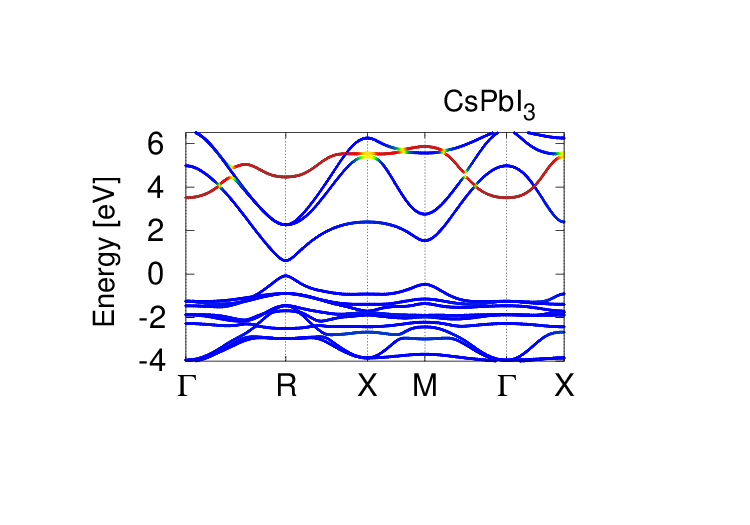} & \hspace{-3.5cm} \includegraphics[scale=0.7,angle=0.0]{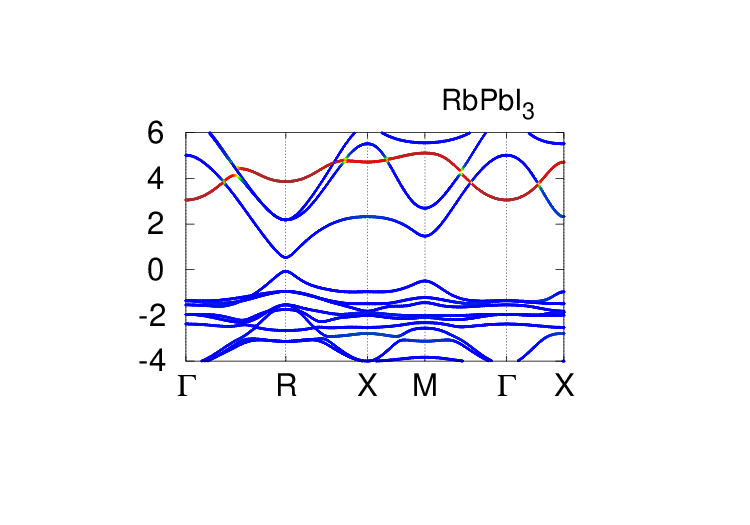} &
\hspace{-3.5cm} \includegraphics[scale=0.7,angle=0.0]{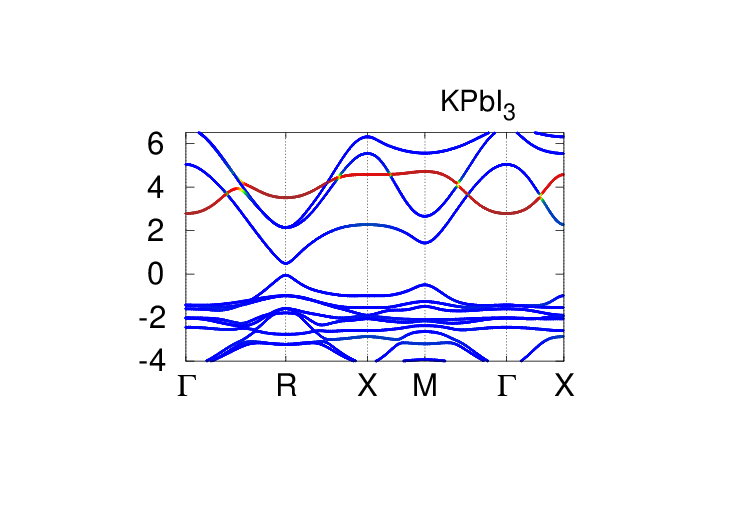} \\[-1.6cm]
\hspace{-1.0cm} \includegraphics[scale=0.7,angle=0.0]{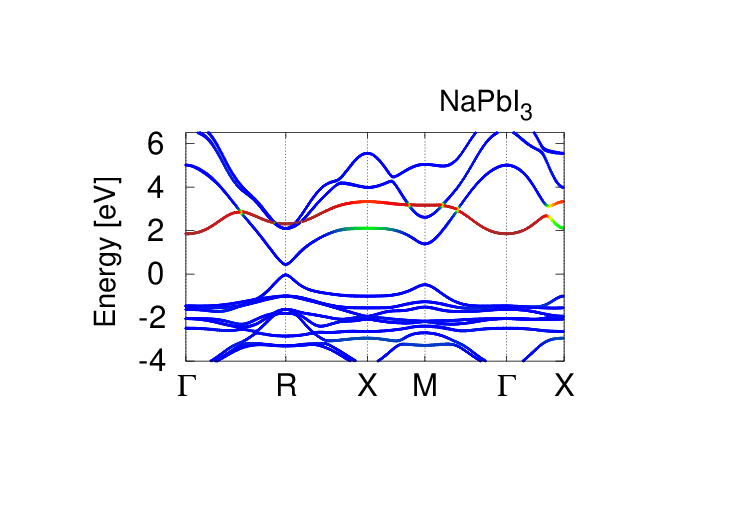} & \hspace{-3.5cm} 
\includegraphics[scale=0.7,angle=0.0]{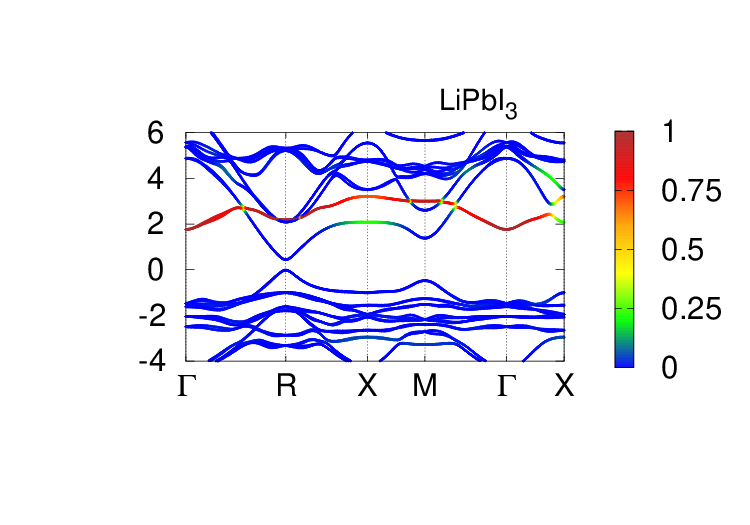} &
\end{tabular}
\caption{GW+SOC band structures of APbI$_3$ perovskites with A = Cs, Rb, K, Na and Li.
Bands projected at maximally-localized Wannier functions (MLWFs) located at the A$^+$ cation empty states are shown in red. 
The legend is common to all cases.}
\label{F1}
\end{figure*}

The band structures obtained within the GW+SOC approximation for the perovskites considered in this work are shown in Fig.~\ref{F1}. 
For completeness, the corresponding bands obtained within DFT and DFT+SOC are included as Figs.~S1 and S2, respectively, 
in the supporting information (SI). In all cases, the band extrema (conduction band minimum, CBM, and valence band maximum, VBM) are at $R$, 
with secondary extrema at $M$. The GW+SOC gaps, band widths and lattice parameters, on the contrary, 
depend on the particular A$^+$ cation, as collected in Table~\ref{T1}. Also for completeness, 
the corresponding values for DFT and DFT+SOC are given in Table S1 in SI. 

In relation to the lattice parameters, the calculated value for CsPbI$_3$ (6.38 \AA) is slightly larger than the experimental value. 
This discrepancy was expectable, since GGA generally overestimates bond lengths. 
The lack of experimental data impedes a proper analysis of our results. 
Nonetheless, it is noticeable that the computed lattice parameters change by roughly 0.12 \AA$\;$ 
for the whole cation series considered herein. This change is smaller than that occurring when iodine 
is replaced by lighter anions: we remark here that the lattice parameters for 
CsPbI$_3$, CsPbBr$_3$ and CsPbCl$_3$ are 5.61 \AA, 5.87 {\AA}  and 6.38 \AA, respectively. 

A relevant feature for the subsequent discussion is the existence of an unoccupied band manifold associated to the A$^+$ cation, 
which may be evinced by projecting the computed bands onto MLWFs centered at each A$^+$ site. 
In Fig.~\ref{F1}, the color code quantifies this projection, with the red (blue) color corresponding 
to the maximum (minimum) value. Our calculations reveal the existence of A$^+$ doublet states within the conduction band in each case. 
As a general conclusion, 
the band gap lowers and the empty A$^+$ cation band shifts down towards the CBM with the decreasing A$^+$ cation mass, 
as reported in Table~\ref{T1}. 
Nevertheless, except at $\Gamma$ (where the bandgap 
is much larger than that at $R$ point), the empty states of A$^+$ cations do not fall below the local CBM. 
The minimum of the A$^+$-band is at $\Gamma$ in all cases, with a secondary one at $R$, 
and its width decreases with the mass of the A$^+$ cation as well. 
We have checked that the lowering of the A$^+$ cation empty states 
has a chemical nature (i.e., it is due to the individual cation in each case), 
and it is not just a geometrical effect (i.e., shorter chemical bonds as the lattice parameter decreases). 
To this end, we have computed the band structure of CsPbI$_3$ using the lattice parameter of LiPbI$_3$,
and {\it vice versa}. The results, shown in SI, Fig.~S3, evidences that the aforementioned lowering is insensitive to changes of the lattice parameter. 
The effect could be related instead to a weakening of the hydrogen bonds between the A$^+$ cations and the free electron pairs of the iodides. 
A fingerprint of this weakening is the fact that the bandwidths of 
the A$^+$ empty states are smaller for the lighter cations. In any case, this issue has not been further investigated herein.

\begin{table}[htb!]
\caption{Characterization of the GW+SOC band structures for APbI$_3$ with A = Cs, Rb, K, Na and Li. 
$E_g^R$ and $E_g^{\Gamma}$ are the bandgaps at $R$ and $\Gamma$, respectively, $\Delta^{CBM-R}$ and $\Delta^{CBM-\Gamma}$ 
are the energy differences between the CBM and the A$^+$-band at $R$ and $\Gamma$, respectively, 
$W_A$ is the width of the A$^+$-band and $a$ is the optimized lattice constant (in \AA). 
All energies are given in eV. }
\begin{tabular}{ccccccc}
\hline \\[-0.2cm]
A & $E_g^R$ & $E_g^{\Gamma}$ & $\Delta^{CBM-R}$ & $\Delta^{CBM-\Gamma}$ & $W_A$ & $a$ \\[0.2cm]
\hline \\[0.1cm]
Cs & 0.81 & 4.40 & $-$4.38 & $-$2.63 & 2.51 & 6.379 \\
Rb & 0.75 & 4.07 & $-$3.04 & $-$2.25 & 2.33 & 6.327 \\
K & 0.71 & 3.89 & $-$2.74 & $-$2.03 & 1.82 & 6.295 \\
Na & 0.67 & 3.03 & $-$1.58 & $-$1.15 & 1.50 & 6.279 \\
Li & 0.46 & 3.25 & $-$1.75 & $-$1.33 & 1.47 & 6.259 \\[0.2cm]
\hline
\end{tabular}
\label{T1}
\end{table}

The chemical character of the A$^+$ cation empty state is the extended s-type function exhibiting 
the interaction with the neighbouring A$^+$ and not the lead halide lattice.
The proof of that is included in Fig.~S4 in SI, where the band structures
and spreads of the MLWFs building the cation empty states are presented for CsPbI$_3$ and the Cs$^+$ cation in two 
periodic lattices: 1) akin to that of the corresponding perovskite ($a$=6.38 $\AA$) and 2) with much larger lattice parameter
($a$=10.58 $\AA$). Important implication of the absence of the hybridization between the A$^+$ cations
and the BX$_3$ part for the optical spectra would be a strong spacial localization of electrons at the highly excited states. 
Such states do not possess the diffuse character that is usual for the conduction states in the ordinary semiconductors.

\subsection{Optical spectra and excitonic behavior}
\begin{figure}[htb!]
\includegraphics[scale=0.35,angle=0.0]{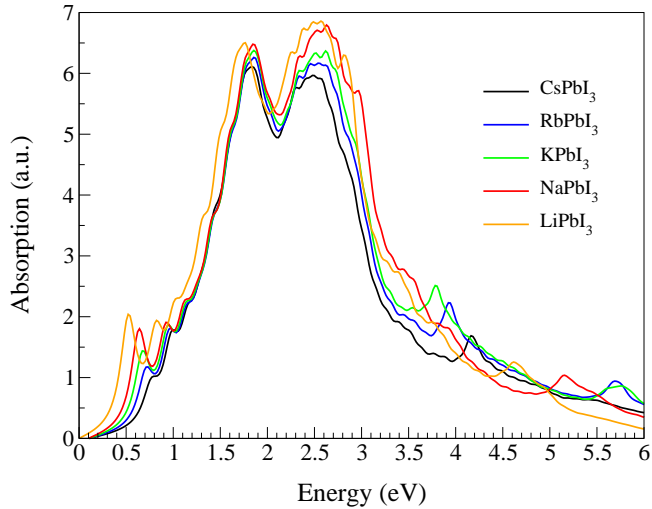}
\caption{Optical absorption spectra calculated with BSE for the perovskites series considered in this work.}
\label{F2}
\end{figure}

The light absorption spectra computed by solving the BSE on top of the GW+SOC calculations for 
the series of perovskites considered herein are shown in Fig.~\ref{F2}. 
The reliability of the calculations has been assessed by checking the convergence of the BSE spectrum with respect 
to the BZ sampling, which is demonstrated in SI (Fig.~S5). 
Regardless of the particular metallic cation, the optical spectra feature two prominent peaks, 
located at similar excitation energy ranges, namely 1.7 - 1.8 eV and 2.6 - 2.8 eV. 
Note that their relative intensities depend upon the choice of A$^+$, though, 
the first peak being more intense only for the heavier cations (Cs$^+$ to K$^+$). 
There is also a low-energy peak at the absorption edge, which decreases in intensity and shifts to 
the blue with increasing the mass of the A$^+$ cation. Above 3.5~eV, less intense but still well pronounced peaks are found in all cases. 
As will be discussed below, the origin of these peaks is different for heavy and light cations. 
For cations from Cs$^+$ to K$^+$, the peaks are related to the optical transitions (and the accompanying charge redistribution) 
from iodide to the A$^+$ empty states. For the lighter cations Na$^+$ and Li$^+$, 
the high-energy peaks in the absorption spectrum correspond to Pb-I transitions. 

The peaks of the absorption spectra, particularly those at the low-energy edges, 
indicate the existence of excitons in the systems. 
Actually, in all cases, the whole spectra up to 6 eV consist of about twenty thousand excitons, 
a number high enough to impede their entire characterization. 
Instead, only  selected optical transitions from low (up to 1.3 eV), middle (1.8-2.8 eV) and high energy (3.3-4.7 eV) parts of 
the absorption spectra have been analyzed.
Particularly, we examined about twenty excitons for all cations, 
representative for the each spectrum up to about 4.7 eV. 
In general, the description of absorption and corresponding emission lines should include  
effects such as polaron formation and strong electron-phonon coupling.\cite{exp1, Rev3} 
These effects are manifested by the temperature dependence 
of the photoluminescence (PL) intensity and the Stokes shift appearing, for example, 
in the triple halide perovskites.\cite{phase_stability} 
However, the effect of phonons\cite{exp1} is beyond our theoretical capabilities, and has not been taken into account in this work.

\begin{figure}[htb!]
\includegraphics[scale=0.4,angle=0.0]{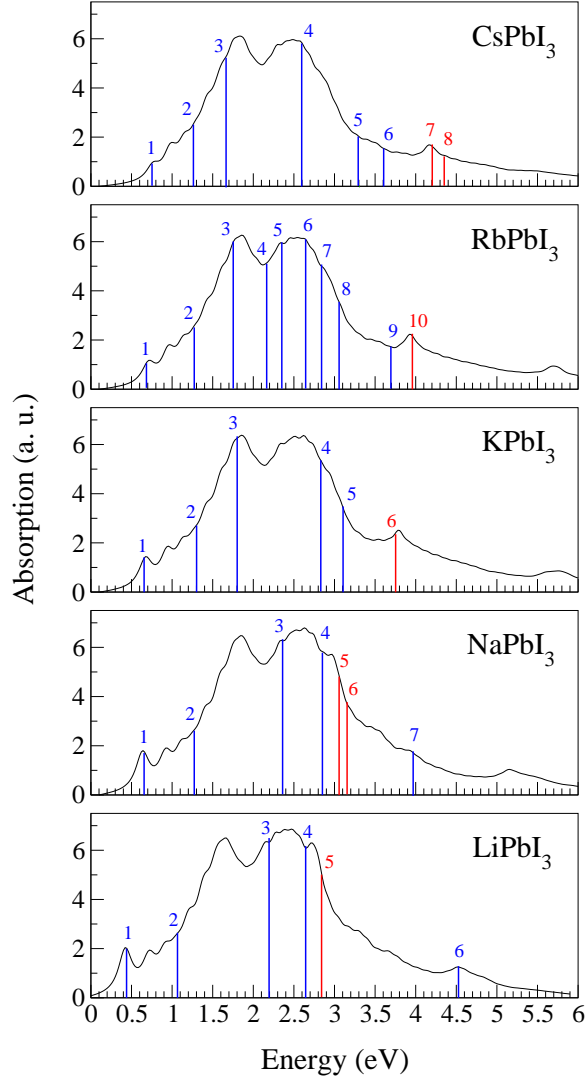}
\caption{Chosen optical dipole transitions in the absorption spectra of APbI$_3$
with A = Cs, Rb, K, Na and Li.
Transitions marked in red correspond to excitons located at the halide atom
and the empty states of A$^+$. The excitonic parameters for the numbered transitions are given in Table~\ref{T2}.}
\label{F3}
\end{figure}

The spectra have been analyzed as follows. 
Each exciton possesses an excitation energy $E_I$ and an oscillator strength, 
and consists of many single dipole transitions (SDTs).
SDTs occur between occupied states ($b_h$, numbered down in energy as H$^{-1,-2,...}$ from the highest occupied (H) state) and
unoccupied ones ($b_e$, numbered up in energy as L$^{1,2,...}$ from the lowest unoccupied (L) state). 
Two $k$-points in BZ are involved for each SDT, those for the initial ($k_i$) and the final states ($k_f$). 
We have chosen herein all SDTs to be direct, and thus $k_i = k_f \equiv k$. On the other hand,  
each SDT may contribute to several excitons with different weights. In this work, we chose a threshold of 5$\%$ for the SDTs weights. 
In other words, for a given exciton, only SDTs contributing with the intensity by more than 5$\%$ are considered. 

The energy of each SDT, $E^{GW}(t)$, where $t$ denotes the set $\{k_i, k_f, b_h, b_e\}$, 
is the difference between the energies of the final and initial states, obtained with the GW method, 
hence excluding any $e-h$ interaction correction. 
We define the  
excitation binding energy ($E_b$) as
\begin{equation}
	E_b(t) = E^{GW}(t) - E^{BSE}(t) \equiv E^{GW}(t) - E_I(t)
	\label{eq:eb}
\end{equation}
where the energy $E^{BSE}(t)$ is obtained with the BSE method and accounts for the $e-h$ interactions. 
For bound excitons (with attractive $e-h$ interactions), the $E^{BSE}(t)$ energies are lower than the interband energies 
$E^{GW}(t)$ and, hence, the values of $E_b$ are positive. 
Note that, since a given SDT (characterized by a set $t$) may contribute to many excitons,
there is not a single $E_I(t)$, but actually a range $[E^{min}(t), E^{max}(t)]$ corresponding to the excitons contributing to the SDT. 
According to Eq.~(\ref{eq:eb}), the highest $E_b(t)$ (that is, the maximum binding energy of the SDT) 
is for the lowest exciton energy that contains the chosen SDT with weight above the threshold, $E_I^{min}(t)$.  
Note that, according to this definition, $E_b(t)$ is mathematically univocal.

We are interested in four characteristics of the excitations: i) their chemical composition, 
which is given by the projection of the involved bands onto atomic states,
ii) their location at high-symmetry $k$-points in BZ, iii) the $E_I^{min}$ values to which they contribute, 
and iv) the value of $E_b$. As was mentioned above, to sample the BSE spectrum we have restricted ourselves to selected 
transitions at low (up to 1.3 eV), middle (1.8\--2.8 eV) and high energies (3.3\--4.7 eV), 
namely those transitions exhibiting the highest oscillator strengths. 
The selected optical transitions are shown in Fig.~\ref{F3} for all cation species, 
where the blue lines are for transitions located at the PbI$_3$ moiety of the crystal, 
and the red ones for those transitions involving the A$^+$ empty states. 
Table~\ref{T2} collects the above four characteristics for these selected SDTs. 
The numbering of the SDTs in Fig.~\ref{F3} corresponds to the $d$ indexes listed in Table~\ref{T2},
and the transitions involving the A$^+$ empty states (in red in Fig.~\ref{F3}) appear in bold in Table~\ref{T2}.

\begin{longtable}[htb!]{ccccc}
	\caption{Excitonic properties of APbI$_3$ with A = Cs, Rb, K, Na and Li: 
chosen SDT ($d$), exciton energy ($E_I^{min}$), 
direct transition $k$-point in BZ, 
bands involved in the dipole transition ($b_h$ $\rightarrow$ $b_e$), 
exciton binding energy ($E_b$). 
	$K^Q$ denotes a small shift from $K$ towards the high-symmetry $k$-point $Q$. 
H$^{-n}$ and L$^{+n}$ denote the highest occupied and lowest unoccupied bands, respectively, at the 
$k$-point of the direct transition, 
and the upper index denote bands below/above these. Energies are given in eV.} \\ 
	\toprule \toprule
	\endfirsthead
	\caption* {Table~\ref{T2} continued}\\ \toprule
	\endhead
	\endfoot
	\bottomrule \bottomrule
	\endlastfoot
	$d$ &	$E_I^{min}$ & $k$-point & $b_h\rightarrow b_e$ & $E_b$ \\ 
\hline 
\multicolumn{5}{l}{CsPbI$_3$} \\
1 & 0.73 & $R$  & H $\rightarrow$ L  & 0.08  \\
2 &  1.28 & $R^M$  &  H $\rightarrow$ L   &  0.11  \\
3 &  1.65 & $R^M$  & H$^{0,-1,-2}$ $\rightarrow$ L &  0.10\--0.15  \\
4 &  2.59 & $M$    &  H$^{-2}$ $\rightarrow$ L   &  0.24  \\
5 &  3.30 & $X$    & H$^{-1}$ $\rightarrow$ L   & 0.16 \\ 
6 &  3.59 & (0,1/8,3/8) &  H $\rightarrow$ L & 0.18 \\
\textbf{7} &  \textbf{4.20} & $\mathbf{\Gamma^{X}\,(\Gamma)}$ & \textbf{H} $\rightarrow$ \textbf{L} & \textbf{0.29 (0.12)} \\
\textbf{8} &  \textbf{4.33} & $\mathbf{\Gamma^{M}}$ & \textbf{H} $\rightarrow$ \textbf{L} & \textbf{0.28} \\
\midrule
\multicolumn{5}{l}{RbPbI$_3$} \\
1 &  0.69 & $R$    &  H $\rightarrow$ L   &  0.06  \\
2 &  1.28 & $R^M$  &  H $\rightarrow$ L   &  0.09  \\
3 &  1.65 & $R^M$  & H $\rightarrow$ L &  0.09  \\
4 &  2.18 & $R$    &  H $\rightarrow$ L$^{+1,+2}$  &  0.13  \\
5 &  2.34 & $M$    & H$^{-1}$ $\rightarrow$ L   & 0.15 \\
6 &  2.65 & $M^R$  & H$^{-2}$ $\rightarrow$ L   & 0.08 \\
7 &  2.81 & (0,1/4,1/2) &  H $\rightarrow$ L & 0.18 \\
8 &  3.08 & $X$    & H $\rightarrow$ L & 0.23 \\
9 &  3.69 & $X^\Gamma$ &  H $\rightarrow$ L & 0.24 \\
\textbf{10} &  \textbf{3.97} & $\mathbf{\Gamma^{X}\,(\Gamma)}$ & \textbf{H}$\mathbf{^{0,-1}} \rightarrow$ \textbf{L} & \textbf{0.28 (0.11)} \\
\midrule
\multicolumn{5}{l}{KPbI$_3$} \\
1 &  0.67 & R    &  H $\rightarrow$ L   &  0.05  \\
2 &  1.29 & R$^M$  &  H $\rightarrow$ L   &  0.07  \\
3 &  1.80 & M$^R$  & H $\rightarrow$ L &  0.13 \\
4 &  2.83 & (1/8,1/4,1/2) &  H $\rightarrow$ L & 0.12 \\
5 &  3.1  & X & H $\rightarrow$ L & 0.21 \\
\textbf{6} &  \textbf{3.85} & $\mathbf{\Gamma^{X}\,(\Gamma)}$ & \textbf{H}$\mathbf{^{0,-1}} \rightarrow $\textbf{L} & \textbf{0.24 (0.08)} \\
\midrule
\multicolumn{5}{l}{NaPbI$_3$} \\
1 &  0.64 & R    &  H $\rightarrow$ L   &  0.03  \\
2 &  1.28 & R$^M$  &  H $\rightarrow$ L   &  0.06  \\
3 &  2.37 & M$^X$  & H $\rightarrow$ L &  0.12 \\
4 &  2.85 & (1/8,1/4,1/2) &  H $\rightarrow$ L & 0.07 \\
\textbf 5 & \textbf{3.06} & $\mathbf{\Gamma^{X}}$ & \textbf H $\rightarrow$ \textbf L & \textbf{0.23} \\
\textbf 6 & \textbf{3.17} & $\mathbf{\Gamma^M}$  & \textbf H $\rightarrow$ \textbf L & \textbf{0.21} \\
7 &  3.97 & M$^X$ & H$^{-1}$ $\rightarrow$ L$^{+1}$ & 0.14 \\
\midrule
\multicolumn{5}{l}{LiPbI$_3$} \\
1 &  0.43   & R    &  H $\rightarrow$ L   &  0.03  \\
2 &  1.08   & R$^M$  &  H $\rightarrow$ L   &  0.06  \\
3 &  2.20  & M & H$^{-1}$ $\rightarrow$ L   &  0.14  \\
4 &  2.62 & M$^{R,X,\Gamma}$ & H$^{0,-1,-2}$ $\rightarrow$ L$^{0,+1}$ & 0.01\--0.1 \\
\textbf 5 & \textbf{2.82} & $\mathbf{\Gamma^X\,(\Gamma)}$ & \textbf H $\rightarrow$ \textbf L & \textbf{0.22 (0.09)}  \\
6 &  4.52   & X    &  H $\rightarrow$ L$^{+2}$ &  0.09  
	\label{T2}
\end{longtable}

For all cations, the highest binding energies of excitons correspond to transitions 
within the ranges 3.8-4.3~eV for heavy cations (Cs$^+$ to K$^+$) 
and 2.8-3.2 for lighter ones, which involve the $A^+$ empty states. 
Very interestingly, these transitions are located not exactly at $\Gamma$, 
but at points shifted towards X or M, which correspond to 
the valence band maxima in this region of the BZ. 
This fact indicates that the hole density of states 
(\textit{i.e.}, the iodide contribution to the transition), affects $E_b$ more than
that of the electron, since the minimum of the A$^+$-derived state is exactly at $\Gamma$ in all cases.
By comparison, the values of $E_b$ computed exactly at the $\Gamma$-point are given in parenthesis in Table~\ref{T2}.

Less strongly bound excitons were observed for transitions from iodine anions to Pb (in blue in Fig.~\ref{F3}); 
these are transitions number 4 for CsPbI$_3$, number 5 for RbPbI$_3$ and number 3 for LiPbI$_3$, all of them located at the $M$ point. 
These transitions correspond to the local valence band maxima and conduction band minima, and yield excitation energies 
within the range 2.20\--2.40 eV. 
High values of $E_b$ were also obtained in some cases for the excitations at the $X$ point, 
where band flattening is observed in Figs.~\ref{F1}. 
These correspond to transitions numbers 5, 8 and 5 for CsPbI$_3$, RbPbI$_3$ and KPbI$_3$, respectively, 
and correlate with excitation energies around 3.1-3.3~eV. 
Finally, the low-energy excitations at the absorption edge are characterized by very low values of $E_b$, 
ranging roughly between 30 and 80~meV, and appear at the $R$ symmetry point.

\subsection{Discussion}
Before analyzing and discussing the previous results, a comparison of these with experimental or theoretical values reported in literature is desirable. 
In this respect, we must admit that performing such a comparison is not a straightforward task. 
The first reason is that $E_b$ values are extremely sensitive to factors such as structure, chemical composition, measurement technique, 
dimensionality, or fabrication routes \cite{phase_stability, exp2, thickness, Wang15, Rev3, Han18}. 
Besides, by contrast to the huge number of papers describing the optical performance of HPs, those focused on IPs are much more scarce. 
Yunakova et al. report the binding energy for CsPbI$_3$ thin films to be $E_b =$ 0.157~meV, but their result is not comparable to ours because 
their crystals were orthorhombic and, besides, chemically impure \cite{exp3}. 
Yang et al., on the other hand, used magneto-transmission measurements to estimate $E_b =$ 15 meV for CsPbI$_3$ stabilized in the cubic phase 
by annealing \cite{exp2}. Their results are admittedly lower than those obtained by magneto-optical techniques, 
which could be due to the fact that the samples are metastable.
 
As for theoretical works, it has been recognized that predicted exciton binding energies are usually well higher 
than experimental ones \cite{exp1, Fuchs08, Bokdam16}. 
In particular, Filip et al. compute $E_b =$ 39 meV by solving the BSE within a dynamic screening approximation \cite{exp1}. 
The value of $E_b$ calculated herein for CsPbI$_3$ at the absorption edge overestimates the results by Filip et al. due to two reasons: 
1) we adopted a static screening approach which completely ignores the screening of the $e-h$ interactions by phonons, and
2) we studied the $\alpha$ cubic phase of CsPbI$_3$ while these authors investigated the orthorhombic phase, among other compounds.
Some comments can be done in this respect. For instance, the bandgaps and absorption edges of lead halide perovskites increase as the symmetry decreases from the cubic to the orthorhombic structures (see Table 3 in our previous work \cite{breath}).  
This fact suggests that the above rule could be also true for  
exciton binding energies at the absorption edges, 
provided that one ignores the phonon screening of the $e-h$ interactions. 
In this sense, it would be very interesting to check the effect of phonons for excitons in different structures, 
as also for excitations at various $k$-points in BZ. The indirect evidence for a large variety of the excitonic
binding energies is given by a measurement of the exciton lifetimes, 
since they are inversely proportional to the $E_b$ values.\cite{Rev3}

It is also interesting to compare the computed and measured absorption spectra. In this case, 
the difficulty arises because experimental data refer mostly to the mixed A$^+$ cation and mixed halide compounds. 
Besides, the high energy region of the absorption spectra has not been explored by optical pumping  
to date (although UV lasers up to 5 eV are available), to our knowledge of the literature. 
Still, one may get additional indirect support to the theoretical prediction of an important role of 
the A$^+$ cation on the optical performance. 
For example, the photostability of (Cs,FA)Pb(I,Br)$_3$ grows very efficiently with a change of the laser pumping energy 
of only 0.1 eV, \cite{optica} 
which is consistent with the highest peaks shifts calculated herein for the corresponding A$^+$ cations and could be an indirect sign of the role played by these. 
Also, increasing the Cs content with simultaneous decreasing that of Pb enhances 
the crystal stability under lasing operation.\cite{triple} 
Finally, recombination centers for excitons are formed when the mixed A$^+$ cations 
segregate to yield an inhomogeneous distribution.\cite{Science} 

As we pointed out in the Introduction, different parts of the absorption spectrum of perovskites may be excited depending on 
their use as component of optical devices. 
In particular, optically-pumped lasers are excited approximately within the range 2.2-3.5~eV, 
for which strongly bound excitons are desirable. 
Our results indicate that such an interval could be realized for suitable choices of 
the A$^+$ cation in lead halide perovskites. 
The corresponding excitons exhibit high binding energies and, importantly, involve A$^+$ empty states. 
In other words, one gets very strongly bound excitons with the electron located at A$^+$ 
which likely yields high power lasing. 
Actually, experimental reports for optically-pumped lasers, based on mixed cation and halide perovskites,
communicate optical gain obtained at the corresponding wavelengths.\cite{optica,R2021} 
However, there are troublesome issues related to crystal deionization under strong laser pumping, 
since excitations from iodine to the A$^+$ cation change I$^-$ and A$^+$ to the neutral pairs I$^0$ and A$^0$, 
which destabilizes the ionic structure of the PbI$_3$ moiety. 
The results reported herein suggest that this problem could be partially solved by designing the mixed cation compounds in such a way, 
that the majority-type A$^+$ cations serve as donors for the PbI$_3$ ionic bonds, 
whereas the minority-type A$^+$ cations add their empty states to be used in excitonic purposes. 
This is a plausibility argument which will be explored further in future works. 

The low-energy edge of the absorption spectrum is rather dominated by weakly bound excitons, 
$E_b$ $\sim$ 30-80 meV. We have already pointed out that the intensity of these peaks grows as the A$^+$ cation becomes lighter. 
Both facts are compatible with the use of APbI$_3$ perovskites as parts of the solar cells, 
and could be promising for the realization of electrically pumped perovskite lasers too,
although this is a highly speculative prediction with the data available to date. 

\section{Summary}
Focusing on 
excitonic properties, 
we have performed \textit{ab initio} MBPT calculations for a series of inorganic lead halide perovskites,
namely APbI$_3$ with A = Cs, Rb, K, Na and Li. The results show a marked effect of the A$^+$ cation on the electronic structure 
and optical performance of these systems. 
The band structures obtained within the DFT, DFT+SOC and GW+SOC formalisms, projected on the MLWFs centered at A$^+$, 
show that the energetic position of the A$^+$ empty states lowers as the cations become lighter. 
On the other hand, the BSE absorption spectra reveal that the peaks at the low-energy edge grow for perovskites 
with lighter cations, Na$^+$ and Li$^+$. 

The low-energy edge of the absorption spectrum is dominated by weakly bound excitons, 
$E_b$ $\sim$ 30-80 meV, values not too high for the realization of solar cells. 
The corresponding transitions take place from the VBM to CBM at $R$. 
Further, the middle part of the spectra contain transitions characterized by more strongly bound excitons 
(E$_b \sim$ 150-200~meV) located at PbI$_3$, without any contribution of the A$^+$ states.
These transitions take place at $M$ and $X$, as well as along the high-symmetry line between the $R$ and $M$ points. 
This energetic region includes the highest of the absorption spectrum and the neighboring saddle point. 
Finally, the higher energy part of the spectrum, from about 2.8 eV (for LiPbI$_3$) to 4.3eV (for CsPbI$_3$), 
is characterized by very strongly bound excitons, with E$_b \sim$ 220-290~meV, 
located at the halide atoms and the empty states of A$^+$. 

\section*{Supporting Information}
Additional data, such as the band structures obtained within DFT and DFT+SOC,
study of the variation of the DFT bands with the lattice parameter and
convergence of the BSE spectrum,
are given in the Supporting Information file. 

\section{Acknowledgments}
The calculations have been supported by the Polish Center of Science,
grant No. 2019/33/B/ST8/02105, and were performed using the Prometheus computer within the
PL-Grid supercomputing infrastructure. 
JJM acknowledges financial support from the Ministry of Science and Innovation (Spanish Government) 
through Grant PID2020-112936GB-I00/AEI/10.13039/501100011033 and by Junta de Extremadura through Grant IB20079, 
both of them partially funded by Fondo Europeo de Desarrollo Regional. 

\bibliography{perovskites}
\bibliographystyle{rsc}

\balance
\end{document}